\documentstyle[12pt]{article}
\newcommand{\be}{\begin{eqnarray}}
\newcommand{\ee}{\end{eqnarray}}

\newcommand{\ve}{\varepsilon}
\newcommand{\as}{\alpha_s}

\newcommand{\kp}{\kappa}
\newcommand{\lm}{\lambda}

\newcommand{\pv}{\vec{p}}

\newcommand{\pl}{|\pv\,|}
\newcommand{\PP}{\mathop{\rm P}\nolimits}
\newcommand{\ra}{\rightarrow}
\newcommand{\LQCD}{\Lambda_{\rm QCD}}
\newcommand{\mT}{m_{\rm T}}
\begin{document}
\pagestyle{myheadings}
\markright{}
\begin{titlepage}
\title{\bf{TEMPERATURE DEPENDENCE OF THE QCD COUPLING}}
\author{M.A.\ van Eijck\\
Institute for Theoretical Physics,\\
University of Amsterdam,\\
Valckenierstraat 65,\\
NL-1018 XE Amsterdam,\\
The Netherlands. \and
C.R.\ Stephens\\
Institute for Theoretical Physics,\\
Rijksuniversiteit Utrecht,\\
Princetonplein 5,\\
NL-3508 TA Utrecht,\\
The Netherlands. \and
Ch.G.\ van Weert\\
Institute for Theoretical Physics,\\
University of Amsterdam,\\
Valckenierstraat 65,\\
NL-1018 XE Amsterdam,\\
The Netherlands.}
\maketitle

\centerline{ITFA-93-11}

\centerline{THU-93/08}

\centerline{hep-th/9308013}
\vskip 0.5truein
\begin{abstract}
We present a one-loop calculation of a gauge invariant QCD beta function.
Using both momentum and temperature renormalization group equations
we investigate the running coupling in the magnetic sector
as a function of temperature and momentum scale.
At fixed momentum scale we find that, in contrast to $\lambda\phi^4$ or QED,
high-temperature QCD is strongly coupled, even after
renormalization group improvement.
However, if the momentum scale is changed simultaneously with temperature
in a specified manner, the coupling decreases.
We also point out in what regime dimensional reduction occurs.
Both the cases $N_f$ smaller and larger than $\frac{11}{2} N_c$
are discussed.
\end{abstract}
\end{titlepage}

\section{Introduction}
One of the most difficult aspects of finite-temperature QCD, and in fact
finite-temperature field theory in general, is the appearance of large
perturbative finite-temperature corrections at temperatures
high compared to any zero-temperature mass scale.
This may be illustrated for the case of $\lm\phi^4$ theory
in 4D. There one finds for the two-point function
\be
\Gamma^{(2)}=m^2\left[1+\frac{\lm T^2}{24m^2}+
O\left(\frac{\lm^2T^3}{m^3}\right)\right],\label{zeroTexp}
\ee
where $\lm$ is the zero-temperature coupling and $m$ is the
zero-temperature mass.
Clearly the above expansion is ill-defined in the regime $T\gg m$.
One possible way out is to perform a partial resummation of the
above as a diagrammatic series, i.e.\ summing the ring or plasmon diagrams.
This is basically equivalent to expanding around a theory of mass
$\mT^2\sim\lm T^2$ rather than a theory of mass $m^2$.
If this is done one finds a non-analytic expansion in $\lm$ \cite{Kapusta}
\be
\Gamma^{(2)}={\lm T^2\over24}\left[1-
\left(\frac{\lm}{8\pi^2}\right)^{1/2}+\cdots\right].\label{finTexp}
\ee

In critical phenomena language the above is akin to looking at a
mean field theory or large-mass limit, the large mass being an induced
finite-temperature mass $\mT$.
Such partial resummations would be inadequate to investigate
the vicinity of a phase transition, such as the deconfinement phase transition
or the electroweak phase transition.
Near a second-order or weakly first-order phase transition one expects
to be expanding around a very small effective-mass theory
as opposed to a large effective-mass theory.
Because in this case $T\gg\mT$ we can regard this regime as being a
high-temperature regime as well.

These problems encountered in finite-temperature field theory arise
because one tries to describe one asymptotic regime in terms of
effective degrees of freedom associated with another.
If one wishes to describe finite-temperature systems, one needs to take
into account that the effective degrees of freedom in the
system are scale and temperature dependent.
Generically there will be (at least) three asymptotic regimes wherein
the physics is qualitatively different.
Again these can be simply illustrated in a $\lm\phi^4$ theory.
The first one is when $T/m\ra\infty$ with $m$ the zero-temperature mass.
In this case one gets driven into a large (finite-temperature) mass regime.
This mass subsequently acts as an IR cutoff thereby eliminating the IR problem.
The second is when $T/\mT\ra0$, $\mT\ra0$.
This corresponds to the massless 4D theory.
The third is when $T/\mT\ra\infty$ and
this corresponds to the massless 3D theory.

There are fixed points of a temperature-dependent renormalization group (RG)
associated with all three asymptotic regimes as shown in \cite{Stephens}.
This corresponds to the fact that in the three different
regimes the effective degrees of freedom that describe the theory are
qualitatively different being infinitely massive 4D, massless 4D or
massless 3D respectively.
If one implements an appropriate temperature dependent RG then the entire
crossover between high- and low-temperature behaviour becomes accessible.
The results show that $\lm\phi^4$ exhibits a second-order
phase transition and that the critical exponents associated with
this phase transition are 3D ones, being associated with
a 3D fixed point, not a 4D one.

In this note we wish to implement this philosophy to
a further understanding of finite temperature QCD.
In section 2 we present the ingredients of a one-loop
calculation, and derive the expression for the $\beta$ function.
In the literature \cite{ANNY} there has been ambiguity to even the sign of
the $\beta$ function due to gauge, vertex and configuration dependence.
Landsman \cite{Landsman} has shown how to get around this by using the
background field method in the Landau gauge which is a special case
of the Vilkovisky-de Witt construction of a gauge-independent
effective action \cite{RebhanVdW}.
In section 3 the momentum and finite-temperature $\beta$ functions
are integrated to give the finite-temperature running coupling.
Finally in section 4 we discuss our results.

\section{Renormalization}
Because Lorentz invariance is broken at finite temperature,
the QCD self energy in the Landau gauge decomposes into
three-dimensionally transverse and longitudinal parts.
Being independent, the transverse and longitudinal parts may behave
differently under the RG, and also different masses
(``electric'' and ``magnetic'' respectively) may be generated.
In the static limit (zero external energy) for a heat bath at rest,
the (inverse) propagator only has non-vanishing time-time and
space-space components.
The transverse part $\Pi^{tr}$ of the self energy
is then found from the trace over the spatial components
\be
\Pi^{tr}(\pl)=\frac{\Pi_{ii}(p_0=0,\pv)}{2\pv\,^2}.
\ee

Owing to gauge invariance, the background field method yields
additional Ward identities that give rise to relations between the
scalar structure functions in the tensor decomposition.
In particular the spatial, static three-point vertex
in the symmetric momentum configuration can be connected to
the transverse self energy and the tree vertex in the following way
\cite{Landsman,Rebhan}
\be
\left.r_k\Gamma^{abc}_{ijk}(\vec{p},\vec{q},\vec{r})\right|_{\rm symm.}=
g\,f^{abc}r_k\left[g_{ij}(p_k-q_k)+{\rm cycl.}\right]
\left[1+\Pi^{\rm tr}(\pl)\right].
\label{threepoint}
\ee
The diagrammatics are specified by the special background-field
QCD Feynman rules of Rebhan \cite{Rebhan}.
These rules are easily extended to finite temperature
\cite{Landsman,Antikainen}.

It is well-known that renormalization of the theory at zero temperature
is sufficient to make Green-functions ultraviolet-finite at all temperatures.
However, we have already mentioned that at high temperatures,
if one uses perturbation theory in terms of zero temperature parameters,
one gets large corrections that make
the finite-order perturbative results unreliable.
Therefore one expects a better description of the temperature dependence
if one expresses the theory at finite temperature $T$ in terms of parameters
characteristic of that temperature.
The essence of this is that when one sets a renormalization condition,
one is deciding to parameterize the physics using the parameter
defined by the condition in question.

We renormalize the theory by requiring
as a renormalization condition that the spatial, static, three-gluon
vertex function is equal to the tree vertex at the momentum scale $\kp$
and temperature $\tau$.
Formula (\ref{threepoint}) then implies that
\be
\Pi^{\rm tr}\left.(\pl,\kp,T,\tau)\right|_{\pl=\kp,T=\tau}=0.\label{RenCond}
\ee
Note that there are now two arbitrary renormalization scales $\kp$ and $\tau$.
The assumption of renormalization point independence
of the bare vertex functions gives rise to a momentum-scale RG
and the finite temperature RG
\cite{FTRG}.
Using eq. (\ref{RenCond}) one finds that the corresponding beta functions
for the effective coupling
$\as(\kp,\tau)=g^2(\kp,\tau)/4\pi^2$ are \cite{Landsman}:
\be
\kp\frac{d}{d\kp}\as(\kp,\tau)=\beta^{(\kp)}(\kp/2\tau)=
\as\,\left.\pl\frac{d\Pi^{\rm tr}}{d\pl}\right|_{\pl=\kp,T=\tau},
\label{betaKappa}\\
\tau\frac{d}{d\tau}\as(\kp,\tau)=\beta^{(\tau)}(\kp/2\tau)=
\as\,\left.\,T\frac{d\Pi^{\rm tr}}{dT}\right|_{\pl=\kp,T=\tau},
\label{betaT}
\ee
where a factor of two in the argument of the $\beta$ function is included
for later convenience.

\section{Results}
In the one-loop self-energy calculation we use the
retarded/advanced formalism developed by Aurenche and Beccherawy \cite{RA},
which gives directly the analytic continuation from the
imaginary Matsubara energies and should in the static case give
the same result as the zero external Matsubara frequency.
Because of the presence of distribution functions in the diagrammatic rules,
conventional methods like dimensional regularization and the use of
Feynman parameters are less convenient than in the zero-temperature case.
For instance in a one-loop integral it is not always possible,
by shifting the integration variable, to remove angular variables
simultaneously from the argument of the distribution function and
from the propagators.
And if we use Feynman parameters, when we change the Feynman parameter
from zero to one, the pole of the integrand can move from below
to above the real axis in the complex energy plane.
We have chosen to regularize the one-loop self-energy integral
in four dimensions by letting the differentiation with respect to
the external momentum act on the integrand of the loop integral and
by keeping the $i\ve$ in the retarded/advanced propagators finite
\cite{Winnipeg}.

After summing the diagrams and performing the angular integrations
the final result for the beta functions can be expressed in terms
of the following one-dimensional integrals:
\be
F_n^\eta(y)&=&\int_0^\infty dx\frac{1}{e^{xy}-\eta}
x^n \log\left|\frac{x+1}{x-1}\right|
\\
G_n^\eta(y)&=&\int_0^\infty dx\frac{1}{e^{xy}-\eta}\PP\frac{x}{(x^2-1)^n},
\ee
where $\PP$ stands for the principal part.
For non-negative $n$ the integrals converge at both ends.
The result is:
\be
\beta^{(\kp)}&=&\beta^B_{th}+\beta^B_{vac}+\beta^F_{th}+\beta^F_{vac},\\
\beta^{(\tau)}&=&-\beta^B_{th}-\beta^F_{th}.
\ee
The bose thermal part and vacuum part are given by ($y=\kp/2\tau$):
\be
\beta^B_{th}(y)&=&N_c\as^2\left[
\frac{\pi^2}{4y^2}-\frac{3}{4}F^1_2(y)-\frac{21}{16}F^1_0(y)+
\frac{25}{8}G^1_1(y)+\frac{7}{4}G^1_2(y)\right],
\label{betaBth}\\
\beta^B_{vac}&=&-\frac{11}{6}N_c\as^2.
\ee
The fermion parts are found to be (if we neglect fermion masses
compared to temperature and momentum scale):
\be
\beta^F_{th}(y)&=&N_f\as^2\left[
\frac{\pi^2}{8y^2}-\frac{3}{4}F^{-1}_2(y)
-\frac{1}{4}F^{-1}_0(y)+G^{-1}_1(y)\right],\\
\beta^F_{vac}&=&\frac{1}{3}N_f\as^2.
\ee
Expansions of the integrands around zero show that these
thermal expressions are infrared convergent.
The distribution functions take care of the ultraviolet convergence
of the thermal parts and also make these parts vanish
in the low-temperature limit.

Derivatives of the functions $F_n^\eta$ and $G_n^\eta$ can be expressed
in terms of these function themselves by performing an integration by parts
\be
\left(y\frac{d}{dy}+2m+1\right)F^\eta_{2m}(y)&=&
2\,G^\eta_1(y)+2\sum_{k=1}^m\frac{1}{y^{2k}}\left[1-(1-\eta)^{1-2k}\right]
\zeta(2k)\,\Gamma(2k),\\
y\frac{d}{dy}G^\eta_1(y)&=&2\,G^\eta_2(y).
\ee
Using these identities we can write the thermal parts
of the beta function as a derivative of a function $f$
\be
\beta^B_{th}+\beta^F_{th}=\as^2y\frac{d}{dy}f(y),
\ee
where
\be
f(y)&=&N_c\left[-\frac{\pi^2}{12y^2}+\frac{1}{4}F^1_2(y)+\frac{21}{16}F^1_0(y)+
\frac{7}{8}G^1_1(y)\right]\nonumber\\
&&+N_f\left[-\frac{\pi^2}{24y^2}+\frac{1}{4}F^{-1}_2(y)+
\frac{1}{4}F^{-1}_0(y)\right].
\ee
This function $f$ vanishes in the low-temperature limit.

The fermion parts are in complete agreement with calculations in
the literature \cite{Antikainen}
(with conventional and with background-field Feynman rules;
the one-loop fermion contribution is the same in both cases).
In the boson term, however, there are small numerical differences
with the results of Antikainen et.\ al.\ \cite{Antikainen}
which may be due to the fact that they
used a background field method in the Feynman gauge.

Integration of the beta functions from a point $(\kp_0,\tau_0)$ in the
$\kp,\tau$-plane with value $\as(\kp_0,\tau_0)$ as initial condition,
then gives the running behaviour of the coupling as a function
of the momentum scale and the arbitrary temperature $\tau$
so that it makes the vertex function
invariant of the renormalization scale up to a wavefunction renormalization.
Thus the ``experimental'' input here, $\as(\kp_0,\tau_0)$, is the
coupling at the temperature $\tau_0$ and at some fiducial
momentum scale $\kp_0$.
 From this one describes all the physics at other scales and
other temperatures.

Integration of the system (\ref{betaKappa}) and (\ref{betaT}) of
coupled differential equations gives:
\be
\as(\kp,\tau)=\left[\as^{-1}(\kp_0,\tau_0)+
\left(\frac{11}{6}N_c-\frac{1}{3}N_f\right)\ln\frac{\kp}{\kp_0}
+f\left(\frac{\kp_0}{2\tau_0}\right)
-f\left(\frac{\kp}{2\tau}\right)\right]^{-1}
\label{alpha}.
\ee
By rewriting the integration constant $\as(\kp_0,\tau_0)$ as
\be
\as^{-1}(\kp_0,\tau_0)=
\left(\frac{11}{6}N_c-\frac{1}{3}N_f\right)
\ln\frac{\kp_0}{\Lambda(\kp_0,\tau_0)}
-f\left(\frac{\kp_0}{2\tau_0}\right),
\ee
where now $\Lambda(\kp_0,\tau_0)$ becomes the arbitrary parameter,
we can bring the above expression for the running coupling (\ref{alpha})
in the form
\be
\as(\kp,\tau)=\left[
\left(\frac{11}{6}N_c-\frac{1}{3}N_f\right)
\ln\frac{\kp}{\Lambda(\kp_0,\tau_0)}
-f\left(\frac{\kp}{2\tau}\right)\right]^{-1}\label{alphal},
\ee
which is useful for comparison with vacuum calculations of
the running coupling constant.
In the zero-temperature limit the function $f$ vanishes,
so that we recover the standard vacuum QCD one-loop result
with $\Lambda(\kp_0,0)=\LQCD$ the usual vacuum QCD scale.
As the renormalization scale $\tau$ is arbitrary,
we can freely choose $\tau$ equal to the physical temperature $T$
and in the next section we will do so.
Similarly the momentum scale $\kp$ is set equal to the physical momentum
$\pl$.

\section{Asymptotic behaviour}
For $y\ra\infty$ we find for the $\beta$ functions
\be
\beta^{(\kp)}\ra-\left(\frac{11}{6}N_c-\frac{1}{3}N_f\right)\as^2
+O\left(\as^3\right),
\ee
which is the standard $\beta$ function for QCD in 4D
with massless fermions, and $\beta^{(\tau)}\ra0$.

In the opposite limit $y\ra0$ we obtain
\be
\beta^B_{th}&=&N_c\as^2
\left[-\frac{21\pi^2}{32y}+\frac{11}{6}+O\left(y^2\right)\right]
+O\left(\as^3\right)\label{betab},\\
\beta^F_{th}&=&N_f\as^2\left[-\frac{1}{3}+O\left(y^2\right)\right]
+O\left(\as^3\right)\label{betaf}.
\ee
Thus, in the high-temperature regime $\tau/\kp\ra\infty$, we have
\be
\beta^{(\kp)}\ra-N_c\as^2\frac{21\pi^2}{16}\frac{\tau}{\kp}
\label{highTBeta}
\ee
which indicates that this $\beta$ function is becoming very large
and negative.
Similarly, as $\tau/\kp\ra\infty$
\be
\beta^{(\tau)}\ra N_c\as^2\frac{21\pi^2}{16}\frac{\tau}{\kp}.
\label{highTbetaT}
\ee
Our sign here is in agreement with that of Landsman \cite{Landsman}.
Unlike Landsman however we do not believe it to be an artefact
of a one-loop approximation.
The sign appears quite naturally as increasing temperature in a
function that depends only on $\tau/\kp$ is equivalent to
decreasing $\kp$, i.e.\ to going into the infrared.

A calculation to two loops of the $\beta$ function is expected
to yield in the high-temperature limit a contribution proportional
to $\as^3/y^2$,
and higher-loop calculations would give higher powers in temperature.
In distinction to perturbation theory without RG improvement
at the actual temperature $T$, increasing powers of $1/y$ in
the high-temperature limit are not necessarily a problem in our RG approach.
This is seen by implementing the change of variables
$h=-\as y\,df / dy$ \cite{JournRel}.
One finds the expansion
\be
\beta(h)=-\varepsilon(y)h-h^2+H(y)h^3+O(h^4),\label{betah}
\ee
where
\be
\varepsilon(y)=-\frac{d^2f}{d(\ln y)^2}\left(\frac{d\,f}{d\ln y}\right)^{-1}.
\ee
The function $H(y)$ will be finite for $0\leq y<\infty$.
The function $\ve(y)$ goes to $1$ as $y\ra0$ and vanishes in the
limit $y\ra\infty$.
This expansion will be valid for all $y$ as long as $h$ is not too large.
The transformation to $h$ is a change of dependent variable in a
differential equation, it cannot change the physics.
It does make manifest, however, the well behaved nature of the perturbative
expansion of the $\beta$ function for $y\ra0$.
In this limit the resultant $\beta$ function is exactly that expected
of zero-temperature three-dimensional QCD.
As $y\ra\infty$ one recovers the $\beta$ function
for zero-temperature four-dimensional QCD.
In conclusion, as long as $h\ll1$, the $\beta$ function expansion,
in particular its sign, is perturbatively reliable.

\section{Discussion}
For the moment we consider the number of fermions small enough to
ensure asymptotic freedom at zero temperature, i.e.
$\frac{11}{6}N_c-\frac{1}{3}N_f>0$.
At fixed, finite temperature, the running coupling (\ref{alphal})
then vanishes in the high-momentum limit in much the same way as at
zero temperature, as an inverse logarithm (see fig.\ 1 where
$N_c=3$ and $N_f=6$).
The solutions of the differential equation show a pole at low momenta, where
the coupling increases beyond the range of reliability of perturbation theory.
Its position is governed by a rational function of the initial condition
when $\kp/T$ is small instead of a logarithmic one as in
the zero-temperature case.
The breakdown of perturbation theory in the infrared is not totally
unexpected as there is no IR cutoff.
No finite-temperature tree level magnetic mass was assumed and no
magnetic mass is induced in one loop (even in RG improved perturbation
theory as is used here).
We note that for the running coupling, the low-momentum limit is not
necessarily the same as the high-temperature limit,
because changing the temperature may result in hopping to another curve
with a different initial condition.
This behaviour is determined by the temperature beta function $\beta^{(\tau)}$.

The temperature-dependence of the coupling at fixed momentum scale is
shown in fig.\ 2.
At zero temperature the coupling approaches a finite value
which can be taken to be the zero-temperature coupling at that scale.
At the high-temperature end the pole shows up again.
The existence of the pole can be seen by using the asymptotic
expression (\ref{highTbetaT}).
However to find its position in terms of zero-temperature parameters
one needs the complete arbitrary-temperature expressions.
The increase with temperature of the running coupling constant
seems to contradict the notion of asymptotic freedom.
However, the standard argument \cite{CollinsPerry} relies on a replacement of
momentum scale by temperature in a vacuum result.
This corresponds to the high-momentum limit discussed above.
In contrast, the coupling constant for growing $T$ at fixed momentum scale
increases.
This is consistent with dimensional reduction \cite{DimRed} which implies
that at high temperature one is driven into the three-dimensional
strong-coupling infrared regime.
In view of (\ref{betah}) dimensional reduction takes place as $y\ra0$,
where $f\ra f_{\rm asymp}:=N_c\,21\pi^2/32y$.
However, this limit can be taken in many different ways.
Also, in order for the result to have any validity the coupling constant
in this limit must not be too large, say $\as<1$.
 From the contour plots in fig.\ 3 one sees that as $T/\kp$ increases,
$\kp$ itself must also increase in order that $\as$ stays
within the perturbative regime.
More specifically, when $f\approx f_{\rm asymp}$, one requires that
$T/\kp$ increase no faster than
$(16/21\pi^2)({11\over6}N_c-{1\over3}N_f)\ln(\kp/\LQCD)$.
Note that our contour plot is different from the one found in
ref.\ \cite{EnqvistKajantie} which shows a decreasing coupling.

With the quark-gluon plasma in mind one particular trajectory
suggests itself because there is, of course, a characteristic momentum scale
for the system given by, for instance, the thermal average $<\!p\!>$
of the momentum.
For  non-interacting, massless quanta in thermal equilibrium $<\!p\!>=3T$.
Following \cite{EnqvistKajantie} we may investigate the behaviour of
the coupling constant along this trajectory.
One finds the coupling to decrease as $1/\ln T$.
However, one should remember that this coupling is small only for
fluctuations in the system which have a certain (albeit typical) momentum.
For $T$ decreasing one enters a strong coupling regime
($\as(\kp=3T,T)>1$) at a temperature which depends on $N_c$ and $N_f$.
For $N_c=3$ and $N_f=6$ perturbation theory breaks down for
$T$ less than $T_b\sim1.8\LQCD$.
Clearly in this region the original assumption of non-interacting
particles is invalid.
One might also consider as in \cite{Enqvist} an ensemble average
of the coupling $\as(T)=<\!\as(\kp,T)\!>$
by integrating over all momenta weighted by a statistical factor.
However, then the problem is that the contributions to the average coupling
from small momenta are too large to be trusted in the confines of
a one-loop calculation, see fig.\ 3.
This point can be taken further to physical quantities in the
magnetic sector of the equilibrium system.
Such a quantity will be computed from an ensemble average over
all fluctuation scales and it will be in the form of an expansion
in $\as(T,p)$.
Hence the contribution to the ensemble average from low-momentum
fluctuations will be large and perturbatively unreliable.

In the above we assumed that $N_f<\frac{11}{2}N_c$,
in other words there are not enough fermions to prevent the
colour charge from being screened at high energies.
If, alternatively, $N_f>\frac{11}{2}N_c$ the theory is no
longer asymptotically free as $\kp\ra\infty$ for fixed $T$.
In fact the theory, at least perturbatively, is now ill-defined in that
limit but is expected to be trivial.
The interesting physics arises by noticing that as $T/\kp$ becomes large,
the fermions, because of the form of the fermionic distribution function,
preferentially decouple from loops relative to the bosons.
This can be seen from the expansions (\ref{betab}) and (\ref{betaf}),
the former being dominant relative to the latter in the $y\ra0$ limit.
Although the fermions decouple from loops, this does not mean that
the high-temperature behaviour is independent of the number of fermions.
The coupling constant $\as(\kp,T)$ is highly sensitive to
the zero temperature initial conditions, i.e. $\as(\kp,0)$
which obviously depends on $N_f$.
Observe that $\beta^{(\kp)}(y)$ has a zero, which we take to be at $y=y_1$ (see
fig.\ 4).
In fig.\ 5 a contour plot is shown.
At fixed, finite temperature, as $\kp$ increases, $\as$ decreases
for $\kp<2Ty_1$ and increases for $\kp>2Ty_1$.
On the other hand, for fixed $\kp$ and along the line $\kp=3T$ the coupling
monotonically increases as a function of $T$.

So what are we to conclude from this one-loop calculation of a
gauge-invariant coupling constant appropriate for the
magnetic sector of QCD?
One might expect that as in the case of $\lm\phi^4$ and
QED one could completely control the high-temperature behaviour.
For instance $\lm\phi^4$ is ``strongly coupled'' in the infrared at
high temperatures but this we can deal with using RG techniques,
since it has a non-trivial three-dimensional IR fixed point.
For QCD however, we have found that for fixed momenta this coupling
has a pole, thereby invalidating a RG-improved one-loop approximation.
The reason for this is, we believe, that QCD is a confining theory ---
we simply do not know how the QCD coupling behaves in the confining regime.
The strong coupling then is a result of dimensional reduction into
the IR confining region of three dimensional QCD.
Without being able to solve the confinement problem
we obviously cannot track the crossover between quark-gluon
degrees of freedom and confining degrees of freedom.
It is quite natural to enquire as to whether this is a one loop artifact,
especially in light of the fact that a magnetic mass is induced at two loops.
Obviously this question cannot be answered without addressing
higher loop effects.

\vskip 8mm
\noindent {\bf Acknowledgements}

\noindent This investigation is financially supported by
de Stichting Fundamenteel Onderzoek der Materie (FOM).
CRS would like to thank Denjoe O'Connor for discussions.

\newpage
\noindent {\bf Figure Captions}

\vspace{0.3cm}
\noindent Fig.\ 1: Plot of solutions of the $\beta^{(\kp)}$
differential equation (\ref{betaKappa}) versus $\kp$ at fixed
temperature for two different values of the integration constant.
$N_c=3$ and $N_f=6$.

\vspace{0.3cm}
\noindent Fig.\ 2: Plot of solutions of the $\beta^{(\tau)}$
differential equation (\ref{betaT}) versus $T$ at fixed
momentum $\kp$ for two different values of the integration constant.
$N_c=3$ and $N_f=6$.

\vspace{0.3cm}
\noindent Fig.\ 3: Contour plot of the running coupling constant
$\as(\kp,\tau)$ (\ref{alphal}) for $\tau=T$ and $\tau_0=0$
where $N_c=3$ and $N_f=6$.
Only below the curve $\as=\infty$ (close to the curve $\as=1000$)
the coupling constant is positive and finite.
Temperature and momentum are measured in units of $\LQCD$.
The dashed line represents $\kp=3T$.

\vspace{0.3cm}
\noindent Fig.\ 4: Graph of the beta function $\beta^{(\kp)}(y)$
for different fermion numbers.
For $N_f > \frac{11}{2}N_c$ the beta function has a zero.

\vspace{0.3cm}
\noindent Fig.\ 5: Contour plot of the running coupling constant
$\as(\kp,\tau)$ (\ref{alphal}) for $\tau=T$ and $\tau_0=0$
where $N_c=3$ and $N_f=24$.
Temperature and momentum are measured in units of $\LQCD$.

\end{document}